# IMPLEMENTING CONTINUOUS HRTF MEASUREMENT IN NEAR-FIELD

*Ee-Leng Tan, Santi Peksi, and Woon-Seng Gan*

School of Electrical Engineering, Nanyang Technological University, Singapore
{etanel, speksi, ewsgan}@ntu.edu.sg

## ABSTRACT

Head-related transfer function (HRTF) is an essential component to create an immersive listening experience over headphones for virtual reality (VR) and augmented reality (AR) applications. Metaverse combines VR and AR to create immersive digital experiences, and users are very likely to interact with virtual objects in the near-field (NF). The HRTFs of such objects are highly individualized and dependent on directions and distances. Hence, a significant number of HRTF measurements at different distances in the NF would be needed. Using conventional static stop-and-go HRTF measurement methods to acquire these measurements would be time-consuming and tedious for human listeners. In this paper, we propose a continuous measurement system targeted for the NF, and efficiently capturing HRTFs in the horizontal plane within 45 secs. Comparative experiments are performed on head and torso simulator (HATS) and human listeners to evaluate system consistency and robustness.

## 1. INTRODUCTION

Head-related transfer function (HRTF) and its time-domain representation, head-related impulse response (HRIR) characterizes the transfer path of sound originating from a certain position in space to the eardrum of a listener. Metaverse incorporates virtual reality (VR) and augmented reality (AR) and creates a platform where people and objects can interact, and many of these experiences would take place in the near-field (NF), especially between one and two arm's length when interacting with humans and objects [1].

HRTF is asymptotically distance invariant as the distance between the sound source and eardrum of the listener increases beyond 1.0m. However, HRTF varies significantly in the NF where distances are less than 1.0m [2]. Diffraction and head shadowing effects result in a low-pass filtering effect of sound sources. Acoustic parallax effects also occur due to the differences in the angle between the sound source and ears with the angle between the sound source and the middle of the listener's head [2]. Near-field HRTFs [3, 4, 5] are measured at various distances to capture important auditory distance cues. Such measurement is time-consuming and the human listener's head must stay relatively still to avoid significant angular errors. This measurement setup can be uncomfortable for human listeners. At the point of writing, numerous far-field HRTF datasets [4, 6, 7, 8, 9, 10] have been developed, and only a few NF HRTF datasets have been developed [11, 12, 13]. In general, NF HRTFs are acquired by three categories of approaches.

The first category of approaches involves synthesizing NF HRTFs from far-field HRTFs. One common technique in this category applies a distance variation function (DVF) on far-field HRTFs to synthesize NF HRTFs [14, 15, 16], which describes the changes of an HRTF corresponding to the distance of the sound source. Distance variation function is based on the formulation of a spherical head model and high-frequency parallax effects are omitted since there are no pinnae in the spherical model.

The second category of approaches synthesizes NF HTRFs using numerical methods, which include the boundary element method (BEM) [17, 18], finite element method (FEM) [19], and finite difference time method (FDM) [20]. These methods usually involve high computational cost and employ non-trivial acquisition processes, such as magnetic resonance imaging (MRI) [18], computed tomography (CT) [18, 20, 21, 22], or 3D scanner [17] to obtain the human listener's geometry.

The third category of approaches performs HRTF measurements in the NF. A set of HRTFs acquired at source distances between 0.25m and 1.5m on a Neumann KU100 dummy head was reported in [23]. However, the sound source used in their experiments might have violated the assumption of an acoustic point source in the proximal region. Another measurement system utilizing 12 loudspeakers mounted on support rods placed in a vertical locating loop with a radius of 1.25m to simulate sound sources at different elevation angles was discussed in [24]. The loudspeakers used in the experiments have diameters of 18mm and have responses similar to a highpass filter with a cutoff frequency at approximately 500Hz.

Our proposed approach falls into the third category. A continuous measurement system [25] has been adapted to efficiently capture NF HRTF in the horizontal plane. At an angular resolution of 5°, the HRTFs in the horizontal plane are captured at an average speed of 45 secs. The performance

This research is supported by the Ministry of Education, Singapore, under its Academic Research Fund Tier 2 (MOE-T2EP20221-0014).

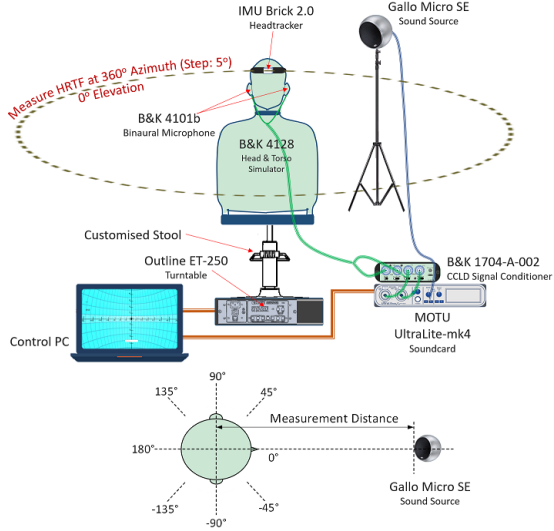

Fig.1. Proposed system for continuous NF HRTF measurement.

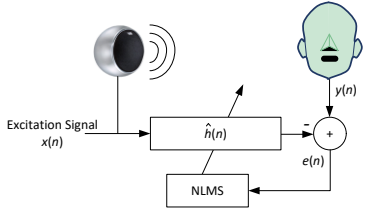

Fig. 2. Block diagram of proposed system.

and consistency of the system are analyzed using spectral difference (SD) and interaural time difference (ITD).

## 2. PROPOSED MEASUREMENT SYSTEM

A block diagram of the proposed system for continuous measurement of NF HRTF is illustrated in Fig. 1. The coordinate system used in our measurement is based on the SOFA standard [26]. The hardware setup of the proposed measurement system comprises the B&K 4128 head and torso simulator (HATS) fitted with the B&K 4101b binaural microphone and IMU Brick 2.0 head tracker. The head tracker and MOTU MK4 soundcard are connected to a control PC over USB, and the Outline ET-250 turntable is connected to the control PC via LAN. The HRTF measurement is managed by the control PC, which synchronizes the playback of the excitation signal on the sound source and recording of the audio signals picked up by the binaural microphones while positioning the listener using the turntable. The binaural microphones are connected to the B&K signal conditioner 1704-A-002 and the output of the signal conditioner is connected to the MOTU soundcard. The Gallo Acoustics micro SE loudspeaker is selected as the sound source for our measurements. The proprietary software managing the measurement system is developed on Unity and is running on Intel Core i7-6700HQ with 16GB of ram at a maximum CPU utilization of 21%.

### 2.1. Measurement Procedure

The HRTF measurements are performed at 360° azimuthal range in steps of 5°. Before each measurement, the HATS is repositioned at the beginning of each measurement. The head tracker is calibrated using a laser pointed to the loudspeaker at an azimuth of 0°, at the ear level of the HATS or the human listener. The binaural microphones are worn on the ears of the HATS and human listeners. The HRTF measurement starts with the sound source emitting white Gaussian noise as the excitation signal and the listener's head is facing an azimuth of 0°. The human listeners are instructed to stay still throughout the HRTF measurement. The head tracker monitors the head movement and pauses the HRTF measurement at a particular azimuth if the yaw or the pitch is larger than an empirically determined threshold.

### 2.2. Selection of Sound Source

Yu *et al.* [27] studied the effects of the size of a sound source on scattering in the NF. In their study, the multipole re-expansion method with a simple model of a spherical head and pulsating spherical sound source is used. Compared to a point source, their simulations revealed that the highest SD attributed to a pulsating source is found at an azimuth of 90° and a distance of 0.15m between the center of the spherical head and sound source. An increment in SD is observed as the distance between the head and sound source decreases, or the size of the sound source increases. Their simulations suggested that the error in NF measurement remains within ±1dB when a sound source having a radius of 0.05m is placed at 0.2m from the origin of the spherical head. Based on the results reported in [27], we have selected the Gallo micro SE as the sound source for our measurement system due to its small diameter at 0.11m while providing wide dispersion of sound at frequencies between 100Hz and 16kHz.

### 2.3. Estimation of HRIR

The measuring process is identical for both left and right ears, and the measurements for the left ear are discussed in this paper. Figure 2 illustrates the block diagram of the proposed system. Assuming the measured HRIR at the listener's ear is a linear-time invariant system, the recorded signal at the ear of the listener can be approximated by

$$y(n) = \mathbf{h}^T(n)\mathbf{x}(n) + v(n), \quad (1)$$

where $\mathbf{h}^T(n)$ denotes the $L$-tap HRIR vector in the current direction $d(n)$, $v(n)$ is the measurement noise, $\mathbf{x}(n)$ is the excitation signal vector which is given as $[x(n) \ x(n-1) \ \cdots \ x(n-L+1)]^T \in \mathbb{R}^L$, $(\cdot)^T$ is the transpose operator, and $n$ is the discrete time index.

The HRIR is estimated by the adaptive filter $\hat{\mathbf{h}}(n)$ and is initialized with zeros. The coefficients of the filter are updated by the normalized least-mean-square (NLMS) algorithm expressed as

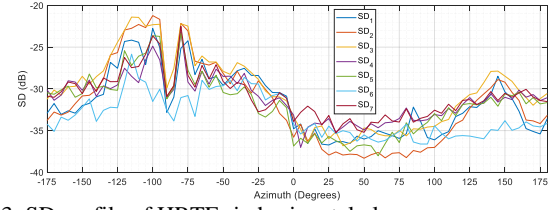
Fig.3. SD profile of HRTFs in horizontal plane.

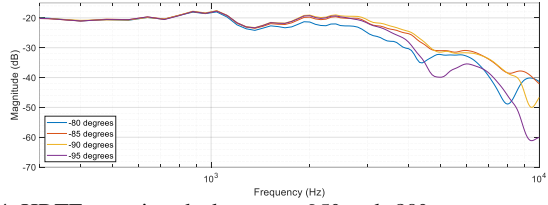
Fig. 4. HRTFs at azimuths between -95° and -80°.

$$\hat{\mathbf{h}}(n+1) = \hat{\mathbf{h}}(n) + \mu(n)\frac{e(n)}{\|\mathbf{x}(n)\|_2^2}\mathbf{x}(n), \quad (2)$$

where the residual error $e(n)$ is

$$e(n) = y(n) - \hat{\mathbf{h}}^T(n)\mathbf{x}(n), \quad (3)$$

with the variable step size scheme [28] expressed as

$$\mu(n) = \begin{cases} \mu_{\max}, & d(n) \neq d(n-1), \\ \max\{\mu(n-1) - \Delta\mu, \mu_{\min}\}, & \text{otherwise.} \end{cases} \quad (4)$$

The above NLMS algorithm resets the step size to $\mu_{\max}$ at the start of each HRIR measurement in a new direction, and gradually reduces to $\mu_{\min}$ at the rate of $\Delta\mu$ in every iteration. In [25], it is found that this scheme speeds up the adaptation of the adaptive filter in each new direction.

As the head of the human listener is unconstrained, there might be several visits in the same direction resulting in multiple HRIR estimations in the same direction. In such scenarios, the lowest normalized mean-square error (NMSE) of an HRIR estimation is selected as the HRIR in a given direction. Considering $J$ visits in the same direction $d$ with these visits ending at iteration indices $n_1, n_2, ..., n_j$, for $i = 1, 2, ..., J$, and $n_i - N_i + 1 \leq n \leq n_i$, where $n_i, N_i$ are the last index and number of samples in direction $d(n_i)$, respectively. The NMSE in direction $d(n_i)$ is computed as

$$NMSE_{d(n_i)} = \frac{\sum_{n=n_i-N_i+1}^{n_i}\left(y(n) - \hat{\mathbf{h}}^T(n_i)\mathbf{x}(n)\right)^2}{\sum_{n=n_i-N_i+1}^{n_i} y^2(n)}, \quad (5)$$

and the filter $\hat{\mathbf{h}}(n_i)$ producing the lowest NMSE is selected as the HRIR in the direction of $d$.

## 3. RESULTS AND DISCUSSIONS

Any movement of a listener during the measurement would affect the consistency and accuracy of the HRTF measurement. To evaluate the consistency of the proposed system, we analyze the HRTF measurements of HATS in the horizontal plane using several performance indices such as SD, normalized correlation coefficient (CC), and ITD. In this section, we focus on distances at 0.4m and 0.6m, which is a good approximation of distances for virtual objects in Metaverse based on the user's arm length [29].

Eight sets of HRTF measurements in the horizontal plane (at resolution of 5°) are performed with the sound source placed 0.4m away from the HATS. The first set of measurements is selected as the reference HRTFs in the horizontal plane, and seven SD profiles are computed for the remaining seven sets of HRTF measurements. To better align SD with human perception [25, 30], the SD values are summed up across 40 warped equivalent rectangular bandwidth (ERB) bands. These SD profiles in the direction $d(n)$ are denoted as $\text{SD}_{q,d(n)}$ for $q = 1, 2, ..., 7$. Based on the frequency response of the sound source, we have selected the 5th to 39th bands in our analysis, and the SD of the $q$th set measurement of HRTF with the reference HRTF in the direction $d(n)$ is

$$\text{SD}_{q,d(n)} = 10\log_{10}\left(\sum_{k=5}^{39} \frac{\sum_{f=f_{c,k}-0.5f_{bw,k}}^{f_{c,k}+0.5f_{bw,k}}\left(H_{ref}(f) - H_q(f)\right)^2}{35\sum_{f=f_{c,k}-0.5f_{bw,k}}^{f_{c,k}+0.5f_{bw,k}} H_{ref}^2(f)}\right), \quad (6)$$

where $H_q(f)$, and $H_{ref}(f)$ are the HRTF of the $q^{\text{th}}$ measurement and reference HRTF in direction $d(n)$, respectively. The center frequency $f_{c,k}$ and bandwidth $f_{bw,k}$ of the $k$th ERB band are

$$f_{c,k} = Q_f BW\left(e^{k\lambda/Q_f} - 1\right), \quad f_{bw,k} = \gamma BW\left(e^{k\lambda/Q_f}\right), \quad (7)$$

where the warp factor $\gamma$, filter $Q$ factor $Q_f$, and minimum bandwidth $BW$ are 1, 9.265, and 24.7, respectively [31].

The SD profiles of the seven sets of HRTF measurements are shown in Fig. 3. Low and high SD values are generally observed near the ipsilateral ears (90°) and near the contralateral ears (-90°), respectively. The higher SD values are primarily caused by the head shadowing effect, and the high complexity of the contralateral HRTFs characterizing the multi-path wave scattering across the head adds up to the difficulty of obtaining accurate measurements at the contralateral ear. This observation of higher SD values at the contralateral ear coincides with the findings reported in [32]. It is interesting to note that a notch (a dip of SD values over 10dB) is observed near the contralateral ear for the SD profiles. To investigate the low SD values at these azimuths, the contralateral HRTFs between azimuths of -85° and -95° are plotted in Fig. 4. It is observed that these HRTFs are fairly smooth within 1kHz and 9kHz. As compared to the HRTFs measured at −95°, the HRTFs measured at -85° and -90° exhibit reduced complexity which in turn leads to lower SD values when compared with the reference HRTF. The contribution of $\text{SD}_{3,d(n)}$ from each ERB band is presented in Fig. 5. The differences between the HRTFs with the reference HRTF mostly lie between the 21st and 39th ERB bands at

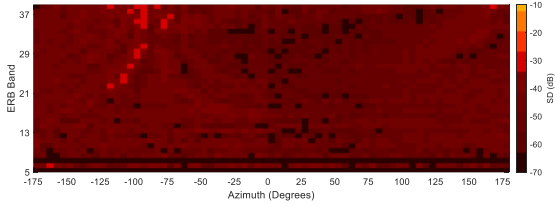
Fig. 5. Values of ERB bands for $SD_{5,d(n)}$.

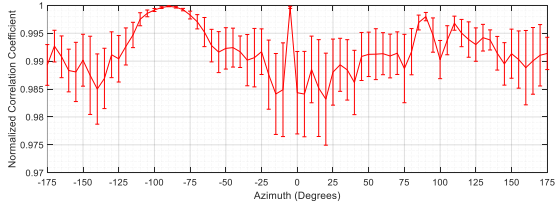
Fig. 6. Mean and standard deviation of ITD (HATS) at 0.4m.

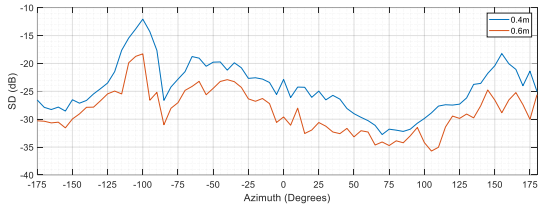
Fig. 7. SD profile of human listeners.

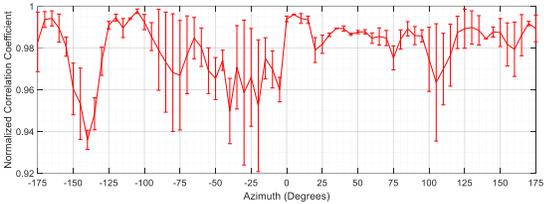
Fig. 8. Mean and standard deviation of ITD (human listeners) at 0.4m.

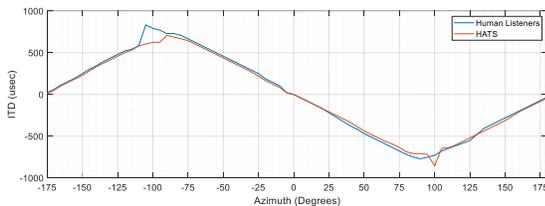
Fig. 9. Mean of ITD measured on HATS and human listeners.

azimuths between -130° and -95°, and 32$^{nd}$ and higher ERB bands at azimuths between -90° and -60.

Next, we compare the consistency of the HRIR measurements using CC. The mean and standard deviation of CC for the seven sets of HRIR measurements with the reference set of measurement at 0.4m is shown in Fig. 6. For all directions, it is found that the CC computed with the seven sets of HRIR measurements are above 0.975, and CC having values near one are mostly located at azimuths approximately 5° and 90°. Referring to Fig. 3, low SD values are also found at these azimuths. For the seven sets of HRIR measurements, the maximum SD value is -21dB and the lowest correlation is 0.975, which suggest that the proposed system has good consistency. It is also noted that the consistency of the system improves slightly with increasing distance. This is expected since the same amount of movement of the listener would introduce less error when the measurement distance is increased.

The average SD profiles (five sets of HRIR measurements each at 0.4m and 0.6m) of four human listeners are plotted in Fig. 7. The SD profile flattens as the distance between the sound source and listener's head increases from 0.4m to 0.6m, and the highest SD values are found at azimuths of -100°. We observe that the SD profiles of the human listeners and HATS exhibit similar trends, but higher SD values are observed for human listeners at all azimuths. The highest SD value is -12dB for the measurements performed at 0.4m, and SD values are reduced for the measurements at 0.6m. The mean and standard deviation of CC between four sets of HRIR measurements and the reference set of HRIR measurement at 0.4m is shown in Fig. 8. The values of CC are found to be above 0.92 at all azimuths.

The interaural difference cues are based on the time difference between the two ears of the listener and any inconsistency in the HRIR measurements can result in localization errors. Let $h_L$ and $h_R$ denote the left HRIR and right HRIR in the direction $d(n)$, respectively. Interaural time difference [33] is given as

$$\text{ITD}_{d(n)} = \arg\max_{\tau} \frac{\sum_{i=1}^{L}\left(h_L(i)-\bar{h}_L\right)\left(h_R(i-\tau)-\bar{h}_R\right)}{f_s\sqrt{\sum_{i=1}^{L}\left(h_L(i)-\bar{h}_L\right)^2 \sum_{i=1}^{L}\left(h_R(i)-\bar{h}_R\right)^2}}, \quad (8)$$

where $\bar{h}_L$ and $\bar{h}_R$ are the means of $h_L$ and $h_R$, respectively, and $f_s$ is the sampling rate. Figure 9 shows the averaged ITD values of HATS and human listeners and the values are found to be similar except for azimuths pointing near the ears.

## 4. CONCLUSIONS

We presented a system that performs continuous HRTF measurement, with a focus on acquiring HRTFs in the NF. The consistency of the proposed system was objectively evaluated using HATs and human listeners. The spectral differences between HRTF measurements in the horizontal plane were quantified using SD, and ITD was used to evaluate the alignment of the ipsilateral and contralateral ears. It was found that the lowest and highest SD values are frequently observed at azimuths pointing towards the ipsilateral and contralateral ears, respectively. Higher contribution of SD values from higher ERB bands and at azimuths pointing towards the contralateral ear were also observed.

Several follow-ups and extensions of this work are planned. First, HRTF measurements of more human listeners are to be performed and these HRTF measurements would form a freely available NF HRTF database. Second, the continuous measurement should be extended to both azimuth and elevation planes for capturing 3D HRTFs. Finally, a MUSHRA test should be conducted for a comprehensive evaluation of the proposed system.

# 5. REFERENCES

[1] Y. K. Dwivedi et al., "Metaverse beyond the hype: Multidisciplinary perspectives on emerging challenges, opportunities, and agenda for research, practice and policy," *Int. J. Inf. Management*, vol. 66, Oct. 2022.

[2] D. S. Brungart and W. M. Rabinowitz, "Auditory localization of nearby sources. Head-related transfer functions," *J. Acoust. Soc. Am.,* vol. 106, no. 3, p. 1465–1479, Aug. 1999.

[3] E. A. G. Shaw and R. Teranishi, "Sound pressure generated in an external-ear replica and real human ears by a nearby point source," *J. Acoust. Soc. Am.,* vol. 44, no. 1, p. 240–249, Jul. 1968.

[4] W. G. Gardner and K. D. Martin, "HRTF measurement of a KEMAR," *J. Acoust. Soc. Am.,* vol. 97, no. 6, p. 3907–3908, Aug. 1995.

[5] P. Majdak, P. Balazs and B. Laback, "Multiple exponential sweep method for fast measurement of head-related transfer functions," *J. Audio Eng. Soc.,* vol. 55, no. 7/8, pp. 623-627, Jul. 2007.

[6] V. R. Algazi, R. O. Duda, D. M. Thompson and C. Avendano, "The CIPIC HRTF database," *presented in Proc. IEEE Workshop on the Applications of Signal Processing to Audio and Acoustics,* pp. 99-102, Oct. 2001.

[7] B. Bernschütz, "A spherical far field HRIR / HRTF compilation of the Neumann KU 100," *presented in Proceedings of the 39th German Annual Conference on Acoustics,* pp. 592-595, Mar. 2013.

[8] K. Watanabe, Y. Iwaya, Y. Suzuki, S. Takane and S. Sato, "Dataset of head-related transfer functions measured with a circular loudspeaker array," *Acoust. Sci. Technol.,* vol. 35, no. 3, pp. 159-165, May 2014.

[9] C. T. Jin, P. Guillon, N. Epain, R. Zolfaghari, A. V. Schaik, A. I. Tew, C. Hetherington and J. Thorpe, "Creating the Sydney York morphological and acoustic recordings of ears database," *IEEE Trans. Multimedia,* vol. 16, no. 1, pp. 37-46, Jan. 2014.

[10] F. Brinkmann, A. Lindau, S. Weinzierl, G. Geissler, S. V. D. Par, L. Aspöck, R. Obdam and M. Vorländer, "The FABIAN head-related transfer function data base," *Technical Report,* Feb. 2017.

[11] T. Qu, Z. Xiao, M. Gong, Y. Huang, X. Li and X. Wu, "Distance-dependent head-related transfer functions measured with high spatial resolution using a spark gap," *IEEE Trans. Audio Speech Lang. Process.,* vol. 17, no. 6, pp. 1124-1132, Sept. 2009.

[12] H. Kayser, S. D. Ewert, J. Anemüller, T. Rohdenburg, V. Hohmann and B. Kollmeier, "Database of multichannel in-ear and behind-the-ear head-related and binaural room impulse responses," *EURASIP Journal on Advances in Signal Processing,* Jul. 2009.

[13] B. Xie, X. Zhong, G. Yu, S. Guan, D. Rao, Z. Liang and C. Zhang, "Report on Research Projects on Head-Related Transfer Functions and Virtual Auditory Displays in China," *J. Audio Eng. Soc.,* vol. 61, no. 5, pp. 314-326, May 2013.

[14] A. Kan, C. Jin and A. V. Schaik, "A Psychophysical Evaluation of Near-Field Head-Related Transfer Functions Synthesized Using a Distance Variation Function," *J. Acoust. Soc. Am.,* vol. 125, no. 4, pp. 2233-2242, Apr. 2009.

[15] A. Kan, C. Jin and A. V. Schaik, "Distance Variation Function for Simulation of Near-Field Virtual Auditory Space," *presented at IEEE International Conference on Acoustics, Speech, and Signal Processing,* May 2006.

[16] S. Spagnol, E. Tavazzi and F. Avanzini, "Distance Rendering and Perception of Nearby Virtual Sound Sources with a Near-Field Filter Model," *Appl. Acoust.,* vol. 115, pp. 61-73, Aug. 2016.

[17] Y. Rui, G. Yu, B. Xie and Y. Liu, "Calculation of Individualized Near-Field Head-Related Transfer Function Database Using Boundary Element Method," *presented at the 134th Convention of the Audio Engineering Society,* May 2013.

[18] C. D. Salvador, S. Sakamoto, J. Treviño and Y. Suzuki, "Dataset of Near-Distance Head-Related Transfer Functions Calculated Using the Boundary Element Method," *presented at the Conference of Spatial Reproduction of the Audio Engineering Society,* Jul. 2018.

[19] F. Ma, J. H. Wu, M. Huang, W. Zhang, W. Hou and C. Bai, "Finite element determination of the head-related transfer function," *J. Mechanics in Medicine and Biology,* vol. 15, no. 5, Apr. 2015.

[20] P. Mokhtari, H. Takemoto, R. Nishimura and H. Kato, "Efficient computation of HRTFs at any distance by FDTD simulation with near to far field transformation," *presented in Autumn Meeting of the Acoustical Society of Japan,* Sept. 2008.

[21] M. Otani, T. Hirahara and S. Ise, "Numerical study on source-distance dependency of head-related transfer functions," *J. Acoust. Soc. Am.,* vol. 125, no. 5, pp. 3253-3261, Jun. 2009.

[22] M. Otani and S. Ise, "Fast calculation system specialized for head-related transfer function based on boundary element method," *J. Acoust. Soc. Am.,* vol. 119, no. 5, pp. 2589-2598, Apr. 2006.

[23] J. M. Arend, A. Neidhardt and C. Pörschmann, "Measurement and perceptual evaluation of a spherical near-field HRTF set," *29th Tonmeistertagung – VDT International Conference,* Nov. 2016.

[24] G. Yu, R. Wu, Y. Lin and B. Xie, "Near-field head-related transfer-function measurement and database of human subjects," *J. Acoust. Soc. Am.,* vol. 143, no. 3, p. EL194–EL198, Mar. 2018.

[25] J. He, R. Ranjan, W. S. Gan, N. K. Chaudhary, D. Hai and R. Gupta, "Fast continuous measurement of HRTFs with unconstrained head movements for 3D audio," *J. Audio Eng. Soc.,* vol. 66, no. 11, pp. 884-900, Nov. 2018.

[26] P. Majdak, Y. Iwaya, T. Carpentier, R. Nicol, M. Parmentier, A. Roginska, Y. Suzuki, K. Watanabe, H. Wierstorf, H. Ziegelwanger and M. Noisternig, "Spatially Oriented Format for Acoustics: A Data Exchange Format Representing Head-Related Transfer Functions," *presented in Proceedings of the 134th AES Convention,* May 2013.

[26] G. Z. Yu, B. S. Xie, and D. Rao, "Effect of sound source scattering on measurement of near-field-related transfer functions," *Chin. Phys. Lett.*, vol. 25, no. 8, Jul. 2002.

[28] G. Enzner, "Analysis and Optimal Control of LMS-Type Adaptive Filtering for Continuous-Azimuth Acquisition of Head Related Impulse Responses," *presented at IEEE International Conference on Acoustics, Speech, and Signal Processing,* Apr. 2008.

[29] T. Edmond et al., "Normal Range of Upper Extremity Length, Circumference, and Rate of Growth in the Pediatric Population," Hand, vol. 15, no. 4, pp. 713-721, Sept. 2020.

[30] R. Nicol, V. Lemaire, A. Bondu and S. Busson, "Looking for a Relevant Similarity Criterion for HRTF Clustering: A Comparative Study," *presented at the 120th Convention of the Audio Engineering Society,* May 2006.

[31] J. Breebaart, F. Nater and A. Kohlrausch, "Spectral and Spatial Parameter Resolution Requirements for Parametric, Filter-Bank-Based HRTF Processing," *J. Audio Eng. Soc.,* vol. 58, no. 3, pp. 126-140, Mar. 2010.

[32] X. L. Zhong and B. S. Xie, "Consistency Among the Head-Related Transfer Functions From Different Measurements," *Proceeding of Meetings on Acoustics,* vol. 19, no. 1, Jan. 2013.

[33] B. F. G. Katz and M. Noisternig, "A Comparative Study of Interaural Time Delay Estimation Methods," *J. Acoust. Soc. Amer.,* vol. 135, no. 6, pp. 3530-3541, Jun. 2014.